\documentstyle[epsf]{l-aa}

\def\la{\mathrel{\hbox{\rlap{\hbox{\lower4pt\hbox{$\sim$}}}\hbox{$<$}}}}
\def\ga{\mathrel{\hbox{\rlap{\hbox{\lower4pt\hbox{$\sim$}}}\hbox{$>$}}}}

\def\arcmin{\hbox{$^\prime$}}
\def\arcsec{\hbox{$^{\prime\prime}$}}

\def\fs{\hbox{$.\!\!^{\rm s}$}}
\def\fdg{\hbox{$.\!\!^\circ$}}
\def\farcm{\hbox{$.\mkern-4mu^\prime$}}

\newcommand{\etal}{{\it et al.}\,}      % et al. in italics
\newcommand{\eg}{{\it e.g.},\ }         % e.g. in italics
\newcommand{\ie}{{\it i.e.},\ }         % i.e. in italics
\newcommand{\cf}{{\it cf.},\ }          % c.f. in italics
\def\deg{{^\circ}}

\newcommand{\Mag}{^m\llap{.\thinspace}}

\newcommand{\kms}{{\,km\ sec^{-1}\,}}

\begin{document}

\thesaurus{03(03.04.1; %Clusters: of galaxies
              07.09.1; %Galaxies: general
              07.19.1: %Galaxies: redshifts of
              09.13.1; % Interstellar medium: extinction
              20.01.2) % Universe: structure of
                      }
\title
{Extragalactic Large-Scale Structures behind the Southern Milky Way.
-- I. Redshifts Obtained at the SAAO in the Hydra/Antlia Extension\thanks
{Table 1 is also available in electronic form. See the
Editorial in A\&AS 1994, Vol. 103, No.1}}

\author{R.C.~Kraan-Korteweg\inst{1} \and A.P.~Fairall\inst{2}
\and C.~Balkowski\inst{3}}

\offprints {Ren\'ee C. Kraan-Korteweg}

\institute{Kapteyn Astronomical Institute,
Postbus 800,
9700 AV Groningen,
The Netherlands
\and
Department of Astronomy, University of Cape Town,
Rondebosch, 7700 South Africa
\and
Observatoire de Paris, DAEC, Unit\'e associ\'ee au CNRS, D0173, et \`a
l'Universit\'e Paris 7, 92195 Meudon Cedex, France}

\date{Received date;accepted date}
\maketitle
\markboth
{R.C.~Kraan-Korteweg et~al.: Large-Scale Structures behind the Southern Milky
Way. - I.}{Kraan-Korteweg et al.}

\begin{abstract}
Spectroscopic observations have been carried out for galaxies
in the Milky Way with the 1.9 m telescope of the South African Astronomical
Observatory (SAAO). The galaxies were selected from a deep optical galaxy
search covering $266\deg \la \ell \la 296\deg, |b| \la 10\deg$
(Kraan-Korteweg 1994). This is in the extension of the Hydra and Antlia
clusters {\em and} in the approximate direction of the dipole anisotropy
in the Cosmic Microwave Background radiation.

The galaxies in the SAAO observing program -- one of the three
complementary approaches in mapping the
3D galaxy distribution in the ZOA -- were selected for high
central surface brightness and for even distribution over the
whole search area. The majority of the galaxies have (absorbed) magnitudes
in the range $14\Mag5 < B_J < 17\Mag5$. Good S/N redshifts
were determined for 115 galaxies. One spectrum
confirmed a planetary nebula, whereas another object with $v=57 \kms$
most likely is galactic as well. The spectra of the other 25 galaxies
either have low S/N or were dominated by superimposed foreground stars.

A preliminary description of the distribution in velocity space
is given. We do find evidence for a continuation of the
Hydra/Antlia supercluster across the ZOA to
$b \approx -10\deg$, making it one of the larger
structures (supercluster?) in the nearby Universe. However, the
prominent overdensity unveiled in the galaxy search
in Vela ($l \approx 280\deg, b\approx +6\deg$) does not, contrary
to what may be expected from the 2D distribution, blend with the Hydra and
Antlia supercluster ($v \approx 3000 \kms$). The latter seems concentrated
at $v \approx 6000 \kms$. Whether it is related to the Great
Attractor cannot yet be assessed. Another weaker concentration is found
at $v\approx 9700 \kms$. Further out, only tentative conclusions
can be made. However, a distinct overabundance of redshifts around
$v \approx 16000 \kms$ is found, corresponding to that of the
adjacent dense Shapley (Alpha) region in
the northern galactic hemisphere {\em and} to that of the
Horologium superclusters in the southern galactic
hemisphere. This might indicate that these two overdensities
are part of a single massive structure, bisected by
the Milky Way.
\end{abstract}


\section{Introduction}

\subsection {Indications of interesting extragalactic structures
in the Southern Milky Way}

Some of the nearest and apparently most influential
concentrations of galaxies in the sky are largely obscured
by the Southern Milky Way ($220\deg<l<360\deg$). Their importance
follows from various lines of evidence.

The dipole of the Cosmic Microwave Background (CMB)
at $\ell=264\deg, b=+48\deg$ (Kogut \etal 1993) implies a motion of our
galaxy of $\sim 600\kms$ towards $\ell=268\deg, b=+27\deg$,
after the Sun's galactocentric velocity is
subtracted and corrections for the motion of our galaxy within
the Local Group (LG) have been applied.
If an infall of about $v_{vc}=220\kms$ towards
the Virgo cluster (Tammann and Sandage, 1985) is taken into account
the direction shifts closer
to the Galactic plane ($\ell=274\deg, b=+11\deg$),
in the region of the Hydra and Antlia clusters
($270\deg, +26\deg$ and $273\deg, +18\deg$, respectively). This leads
immediately to the question as to whether unknow extragalactic
features -- which might prove relevant in explaining the dipole
motion -- are hidden behind the obscuration layer of our Milky Way.

Such a previously unsuspected overdensity has, for instance,
been predicted in the zone of avoidance (ZOA) in Puppis ($l=245\deg$),
based on an analysis of IRAS galaxies using spherical harmonics by
Scharf \etal (1992). This overdensity seems to match successfully a
nearby cluster identified by one of us through HI-observations of
obscured galaxies (Kraan-Korteweg and Huchtmeier 1992). From a subsequent
analysis in this area, we concluded that this previously
unknown cluster is second only to Virgo in the nearby Universe and
adds at least $30\kms$ to the motion of the LG
perpendicular to the SGP (Lahav \etal 1993).

The peculiar motions of galaxies can be established by comparing
redshifts and distances estimated by the Tully-Fisher method
for spirals (Aaronson \etal 1982) and the modified Faber-Jackson
relation for ellipticals (Lynden-Bell \etal 1988). Such peculiar motions
have revealed large-scale streaming motions towards a "Great
Attractor" in the southern sky, close to the galactic plane.
An update by Dressler (1991) predicts its center
at $\ell=307\deg, b=+9\deg, cz=4350 \kms$, while the most recent
density reconstructions from the analysis of the assumed irrotational
peculiar velocity field (POTENT) put it even closer to the galactic
equator at $\ell \approx320\deg, b\approx0\deg$
(Kolatt \etal 1994).

The existence of this massive overdensity is
still quite controversial. Dressler and Faber (1990a,b)
find weak evidence for backside infall into the putative Great Attractor,
but Mathewson \etal (1992) and Hudson
(1993,1994) dispute this and claim that no major mass density can be hidden
in the ZOA. They, and others, using deep IRAS-redshift
surveys and cluster samples (\eg Rowan-Robinson \etal 1990,
Scaramella  1991, Lauer and Postman, 1993, Plionis \etal 1993),
suggest
a residual bulk motion caused by much more distant overdensities
($v > 10000 \kms$). But this is contradicted with investigations by
\eg Lynden-Bell \etal (1989), Strauss \etal (1992), and
Jerjen and Tammann (1993), who find convergence of the dipole motion
between $3500 \kms$ to $6500 \kms$. Still, it should be noted that this
large-scale bulk motion -- if real -- is also located in the general
vicinity of the CMB dipole and the southern Milky Way.

Whatever the case, the mass distribution hidden by the Galactic
zone of avoidance appears crucial. This is manifested also
with the recent POTENT analysis by Kolatt, Dekel and Lahav (1994), in
which they show that the gravitational acceleration at the Local Group
changes by $31\deg$ when the matter distribution from within
$|b|<20\deg$ is included, bringing it very close to the CMB dipole.

If mass overdensities are visible as galaxy overdensities (\eg
Dekel, 1994 \S6.1)
it is important to unveil the distribution
of galaxies behind the southern Milky Way - with particular
interest in the CMB area ($\ell\approx275\deg$) and the Great
Attractor region ($\ell\approx315\deg$). Such is the purpose of
an optical galaxy search started by one of us. It concerns a region of
the ZOA, with $266\deg<\ell<296\deg, |b|\la10\deg$ (cf. Kraan-Korteweg
1989,1992a,b,1994), \ie in the close proximity of the CMB dipole. At the
time of writing, this survey is being continued in collaboration with
Woudt beyond $\ell=296\deg$ to $\ell \la 330\deg$ to cover
the "Great Attractor" region (cf. Kraan-Korteweg and Woudt 1994b, for a
status report on these surveys).

In order to map the
filamentary features and overdensities discovered in
3-dimensional space we have started to measure the redshifts of a
representable part of these galaxies.
This paper reports on redshift observations made with the 1.9m telescope
of the SAAO in the ZOA in the extension of the Hydra and Antlia clusters.

The optical search procedure and the detected galaxy distribution
in the Hydra/Antlia region is reviewed in section 1.2. In
Section 1.3, our observing strategy in determining the 3-dimensional
distribution of these galaxies is outlined. Section 2 then concentrates
on the optical spectroscopy of the brighter galaxies in our search
area that was done at the SAAO:
the observing and reduction procedures are
discussed in 2.1, and the resulting recession velocities
together with the optical parameters of the observed
galaxies are presented in 2.2. This is followed in section 2.3
with a discussion of the properties of the observed galaxy sample,
including velocity data for 31 galaxies in
our sample region taken from other sources. In the third section
we describe the resulting galaxy distribution in redshift
space, and discuss the indicated clustering and filamentary
structures, in context to known large-scale features
adjacent to the zone of avoidance.

\subsection {The galaxy search behind the Milky Way in the extension
of the Hydra/Antlia clusters}

The obscuration of the Milky Way reduces the apparent
magnitudes and apparent angular diameters of galaxies (cf. Cameron 1990).
For this reason, galaxies which lie close to the galactic plane fail to
meet the criteria for inclusion in catalogues (such as \eg Lauberts
1982) and only few galaxies are known below galactic latitudes of
$|b| \la 10\deg$. Added to this are the enormous numbers of foreground
stars that frequently fall on the galaxy images and crowd the
field of view. A separation of galaxy and star images
cannot as yet be done by automated measuring machines such
as COSMOS or APM on a viable basis below $|b| \la 10-15 \deg$.
Examination by eye is still the best technique, though
surveys by eye are clearly both very trying and time
consuming and maybe not as objective.

%\special{psfile=fig1.ps}
\begin{figure*}
\hfil\epsfbox{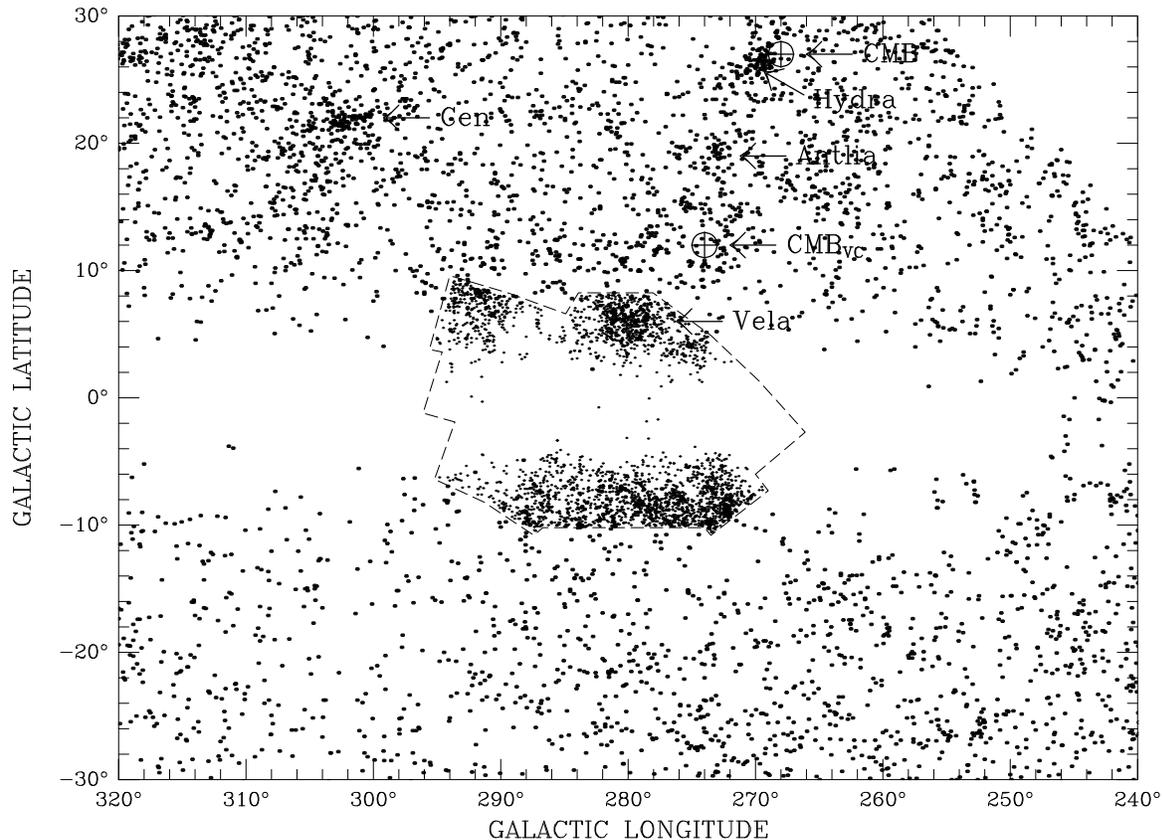}\hfil
%\vspace{11cm}
\caption
{Distribution of galaxies in the Hydra/Antlia extension.
The outlined area marks the area searched in the ZOA. The
3279 galaxies (D$\ga0\farcm2$) unveiled there are shown as small dots.
In the surrounding area the Lauberts galaxies are displayed (large dots,
D$\ge1\farcm$0). The Hydra, Antlia, Vela and Centaurus clusters are
identified, and the CMB-dipole direction, observed (CMB) as well as
corrected for virgocentric infall (CMB$_{VC}$ with $v_{VC} = 220\kms$).
}
\label{f1}
\end{figure*}

Nevertheless, such a survey has been carried within our region of
interest. The tools for this galaxy search are very simple. It comprises
a viewer with the ability to magnify 50 times and the
IIIaJ film copies of the ESO/SERC survey. Full details about
this galaxy search will be presented in the galaxy catalogue
which is in preparation (Kraan-Korteweg 1994 [KK94]).
The success of the procedure for identifying extragalactic objects
at very low latitudes is underlined in the present survey, where of the
142 candidates, only three proved to be foreground stars -- or were
obscured by foreground stars -- and only two were found to be galactic
nebulae.
The main characteristics of this search and the typical
properties of the discovered galaxies have been discussed already
in earlier work (\eg Kraan-Korteweg, 1989, 1991, 1992a,b,
Kraan-Korteweg and Woudt 1994b) and only the most
important aspects will be reviewed here briefly.

The galaxy search in the ZOA in the Hydra/Antlia extension
covers the area $266^\circ \la \ell \la 296^\circ$ and
$-10^\circ \la b \la +8^\circ$. This region of approximately
400$\Box^{\circ}$ encompasses 18
fields of the ESO/SERC survey (F91-F93, F125-F129, F165-F170, F211-F214)
and its boundary is outlined in Fig.~1.

A diameter limit of $D \ga 0\farcm2$ was imposed. Below this diameter
the reflection crosses of the stars disappear, making it hard to
to differentiate consistently between stars or blended stars
and faint galaxies.

In this way 3279 galaxies have been discovered of which only
97 galaxies were previously recorded by Lauberts (1982).
The galaxies found in our search are entered as small dots
in Fig.~1. It is obvious from the distribution of these newly
identified galaxies that the band of obscuration has been
narrowed down to $-4\deg \la b \la 1\fdg5$. There are even a
handfull of galaxies within those limits, suggestive of
thinning of the dust layer at $\ell\approx278\deg$ and $\ell\approx293\deg$.
The variation in galaxy density, however, reflects more than
obscuration by the Milky Way. There are conspicuous overdensities,
foremost around $\ell=280\deg, b=+6\deg$ (the Vela overdensity),
$\ell=275\deg, b=-9\deg$ and $\ell=292\deg, b=+8\deg$,
indicative of extragalactic large-scale structures present in the
survey area.

In order to view the connectivity of these features with their surroundings
we have added the galaxies from the ESO Catalogue (Lauberts 1982) to
Fig.~1, \ie all known galaxies with $D \ge 1arcmin$. The most interesting
features are marked, including the CMB dipole direction.
The concentration of these galaxies in the top left corner
is part of the Centaurus Supercluster with $cz=4500\kms$ (\eg Fairall
and Jones 1991). Our local (Virgo) Supercluster
is an appendage to this dominant structure. At the top, just to the right
of the center, a conspicuous filamentary structure extends down towards
the Galactic Plane. It includes Hydra ($cz=3000\kms,
\ell=270\deg,b=+26\deg$), Antlia ($cz=2800\kms, \ell=273\deg, b=18\deg$),
and the Vela overdensity unveiled in our search.
Most of this supercluster complex is visible
in Fig.~1. There is also an extension from Centaurus to Pavo
($cz=4500\kms, \ell=330\deg, b=-25\deg$).

It is obvious from the
distribution of ESO-galaxies that a significant portion of the supercluster
complex is hidden by the Milky Way but that we can trace these structures
much closer to the galactic dust equator with our deep
galaxy search. However, the measurement of radial velocities is
required to (a) confirm the extragalactic origin of
of these overdensities, (b) to
map them in three dimensions, and (c) to link them to extragalactic
large-scale features above and below the Zone of Avoidance.

\subsection {Ongoing observational programs to outline the 3-dimensional
galaxy distribution}

In collaboration with different groups, various
observational programs were started with this goal in mind.
The three ongoing approaches so far are multifiber
spectroscopy in high density areas, optical spectroscopy
of individual galaxies with high central surface brightness (HSB),
and HI-observations of low surface brightness (LSB) extended
spiral galaxies. The different approaches are complementary with
regard to the results they can achieve and the structures they will
trace (cf. Kraan-Korteweg \etal, 1994a for a discussion on this).

The multifiber spectroscopy is optimal for the high-density areas.
These observations are being done
at the 3.6m telescope of ESO with Optopus and Mefos. The results are
in preparation and will be presented in paper II and III of these
series by Cayatte, Balkowski and
Kraan-Korteweg. We find that the multifiber spectroscopy -- as expected --
preferentially picks up condensed clusters. They furthermore allow
a glimpse at the galaxy distribution through selected small fields
(windows) in the ZOA out to large distances ($v \la 25000 \kms$).
These observations are therefore also important in
outlining structures on very large scales (walls, sheets, etc.).

For determinations of the peculiar velocity of the Local Group (LG) and
the mapping of the velocity flow fields, a redshift
coverage of the nearby galaxy population -- as complete as possible --
is essential. This is not
straightforward to realise. Because of the thick foreground layer
of gas and dust we are seeing only diminished images of the galaxies.
Our minimal goal was to observe all the 'brighter' and 'larger'
galaxies within our search area, and, in addition,
get a fairly homogeneous coverage over the whole search area.

This can be achieved with optical spectroscopy of individual galaxies with
a fairly high central surface brightness, i.e. early type galaxies
or spirals with a distinct bright bulge. The extremely low surface brightness
(LSB) spiral galaxies are mainly late-type galaxies and irregulars, \ie
gas-rich, and can be detected in HI. Together, the methods are well suited
in tracing nearby filamentary structures, and complementary in the sense that
the optical observations allow redshift determinations of early-
type galaxies (which cannot be observed in HI) whilst the 21cm line
observations -- unaffected by foreground obscuration -- allow
the detection of the most absorbed LSB spiral galaxies close to the
galactic dust equator (which cannot be observed in the optical).

Both methods are being undertaken. The program to observe LSB spiral
galaxies is being pursued with the 64m radio telescope at Parkes by
Kraan-Korteweg, Henning, Schr\"oder and van Woerden and will be published
as paper IV of these series.

Meanwhile, three weeks of observing time with the 1.9m telescope of the SAAO
have been devoted to the optical spectroscopy of obscured
galaxies, with relatively high central surface brightness. In the allocated
period, we observed most galaxies in the survey region from which a reasonable
signal-to-noise ratio spectrum could be expected within one hour
of exposure time.

\section {Observations at the SAAO of Galaxies Selected in the ZOA}

\subsection {Observing and Reduction Procedures}

The spectroscopic observations were made in March and April 1991,
and in January 1992, at the South African Astronomical
Observatory (SAAO) at Sutherland. We used the 1.9m Radcliffe reflector,
with "Unit" spectrograph and reticon photon-counting
detector (Jordon \etal 1982). The detector has two
one-dimensional arrays, each 1872 pixels, set to monitor
positions $30\arcsec$ apart in the sky: thus one array
records the galaxy plus sky while the other measures only the sky. To
improve sky subtraction, galaxies were observed on each of
the arrays in turn - unless close neighbouring stars made
this impossible. All galaxy integrations were
bracketed by argon lamp arcs. Maximum time between arcs was
$1500 sec$ with the telescope near the meridian, and less
when aimed well off the meridian and subject to possible
flexure in the instrument. In general, the brighter galaxies
required only one pair of integrations. For many of the
fainter galaxies, multiple pairs of integrations were necessary
to yield reasonable S/N spectra, the total integration time sometimes
exceeding an hour. The
observing routine included exposures of radial-velocity
standard stars (mainly during twilight) as well as
"flat-fields" at the beginning and end of
each night. The dispersion used was approximately $210 \AA/mm$
(Grating 7), corresponding to $2.8 \AA$ per pixel. The slit
width ($300\mu m$) corresponded to $1.8\arcsec$ and the
length of each slit segment, in front of each array, to $6
\arcsec$. Wavelength coverage was $3500-7000 \AA$, with peak
sensitivity in the blue.

Normal reduction procedures have been carried out, using
software written by D. O'Donoghue and J. Menzies, on the
University of Cape Town Vax 6230 computer. Calibration arcs
were fitted with $5^{th}$ order polynomials. Various templates
for the cross-correlation measurement of absorption line
redshifts were examined. The best proved to be a template made
from a mixture of several G and K-type radial-velocity
standard stars. Various checks (\eg manual versus
 cross-correlation, reduction of calibration arc as
 if it were a
galaxy) have been carried out to ensure that there were no
hidden errors in the software.

Due to the faint magnitudes of the target galaxies, we opted for
a low dispersion in the spectrograph: the pixel width (near maximum
instrumental sensitivity) corresponds to about $200\kms$ in
radial velocity. True external errors - standard deviations
and systematic shifts - can be established using the
observations of the radial-velocity standard stars. When the
cross-correlation template was run against the eight
individual stars that comprise it, a standard deviation of
only $18\kms$ was found. However, a more realistic measure
results when the radial velocity standards, observed
in the three observing periods, were run against each other. In
this way a standard deviation
of $35\kms$ was determined, with cross-correlation contrast factors of
$r > 10$.
This value can be taken as the minimum external error for
the system, as used in this observing program.

If the galaxies were bright and
their peak counts per channel comparable to those of the the radial
velocity stars ($>500$), the errors would be of this order.
Obviously, this is not the case. The errors listed in the
accompanying table are derived from r contrast factors, with
the constraint that the error tends down to $35\kms$ as $r > 10$
and rises to $250\kms$ for a marginal contrast of $r = 2.0$. The
$250\kms$ is established from comparing velocities, previously
derived with the same instrumentation, to those from other
investigators (see below). No significant systematic errors
were found, but a small offset ($28\kms$) was applied to the
1992 run as indicated by the standard radial-velocity stars.
The cross-correlation is not used blind; manual checks are
made for each galaxy. For faint marginal galaxies, the
cross-correlation is sometimes replaced by manual
measurements, with appropriate estimated errors. The aim is
always to provide a true external error. One of us (AF) has
found that true external errors are typically
twice the errors claimed by other authors.
This became quite obvious while cataloguing galaxy redshifts
(Fairall and Jones, 1991) and comparing multiple
 radial-velocity
measurements and their quoted errors for a given galaxy.

In the case of emission lines, an error of $100\kms$ has been
assumed for a single line. Thus the error is $70\kms$ when two lines
are used for determining an average, $58\kms$ when three
lines are used, etc.

\subsection {Results}

In this way we have observed 142 objects in the three weeks of
observing time allocated to this program in 1991 and 1992.
The results based on the observations at the SAAO are presented
as follows: we first concentrate on the
galaxies for which a good S/N redshift could be deduced
(2.2.1). This is then compared to other redshift determinations in 2.2.2.
The modest 1.9m aperture of the telescope meant that only
the brighter end of the apparent luminosity distribution
could be observed. Even so, about 20\% of the observed
galaxies proved too faint to extract a reliable redshift
from the spectrum. They and other galaxies for which no reliable
redshift could be obtained are described in 2.2.3.
As discussed before, we aimed at getting redshifts
for all the brighter galaxies in our search area. The
largest of the galaxies in the ZOA (D$\ga 1 arcmin$) were known before
our deep galaxy survey and some of them have already been observed.
As these galaxies are relevant to our investigation they are listed
2.2.4. In the next section (2.2.5) we then turn to a discussion
of the typical properties of the galaxies observed in the ZOA.

\subsubsection {Galaxies with redshifts from the SAAO observations}
In Table~1, the 116 galaxies for which a reliable velocity could
be deduced are listed.
The positions and the most relevant properties,
such as large and small diameters, magnitudes
and morphological types, are taken from KK94 and a detailed discussion
of these parameters is given in the catalogue there. It must be emphasized
again that the observed properties
such as diameters and magnitudes are severely reduced by the
dust screen of the Milky Way. Even at
the highest latitudes ($|b| \approx 8-10\deg$)
the galaxies suffer a foreground extinction of about 1 mag,
decreasing \eg the visible diameter to $\approx 75\%$. The
extinction increases dramatically as we approach the galactic dust equator.
The galaxies in the deepest extinction layer -- which are
still barely visible -- are seen through an obscuration layer of
$A_B \approx 5\Mag0 - 6\Mag0$. Their true diameters will be
a factor $\approx$ 8-10 larger than seen on the IIIaJ plates.
The properties listed in Table~1 are the {\it observed} properties
uncorrected for extinction (and always are underestimates).
%\special{psfile=table1a.ps}
\begin{table*}
\caption{Galaxies with radial velocities obtained at the SAAO}
\hfil\epsfbox{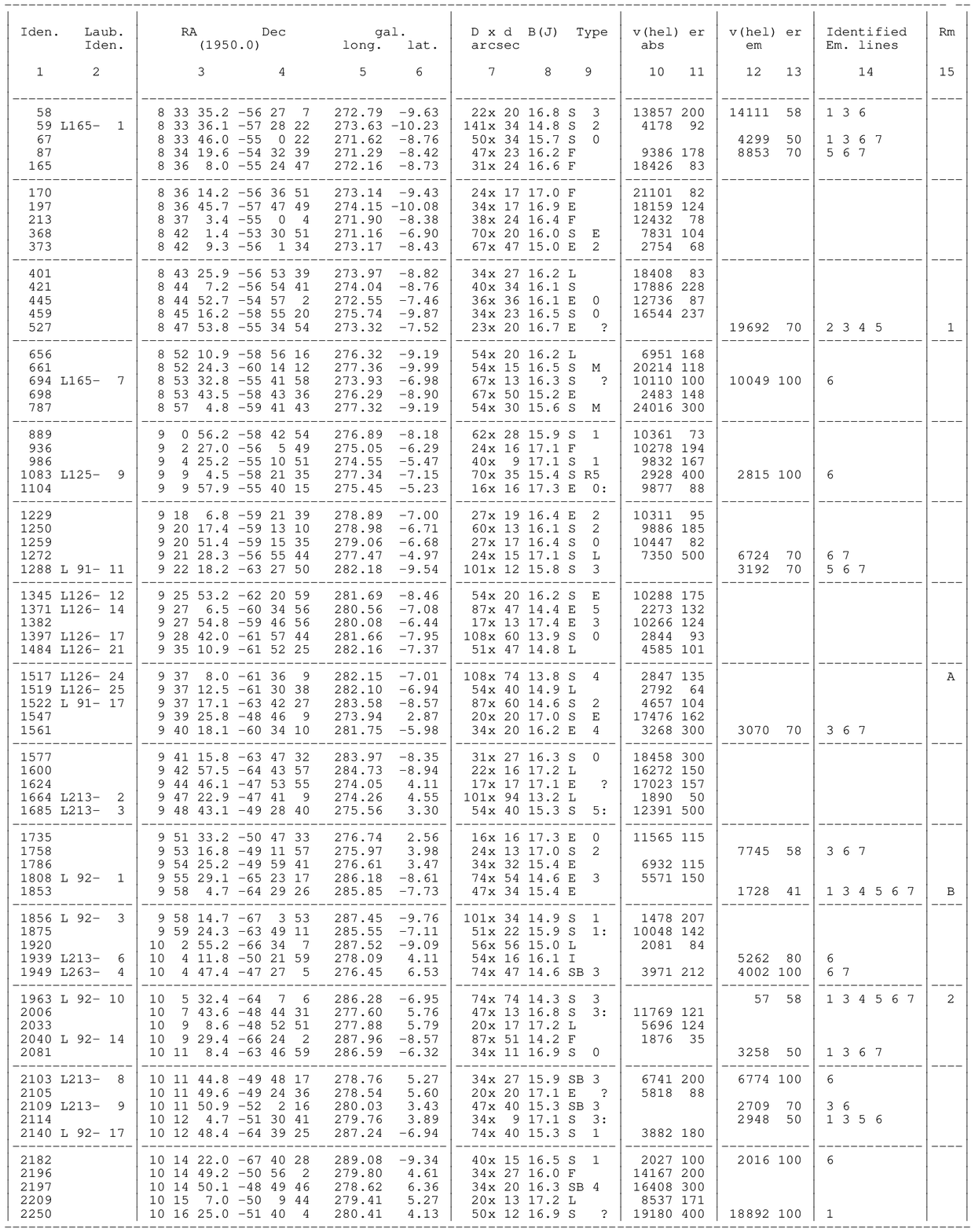}\hfil
%\vspace{21.5cm}
\label{t1}
\end{table*}
\addtocounter{table}{-1}

%\special{psfile=table1b.ps}
\begin{table*}
\caption{-- continued}
\hfil\epsfxsize 1.0\textwidth \epsfbox{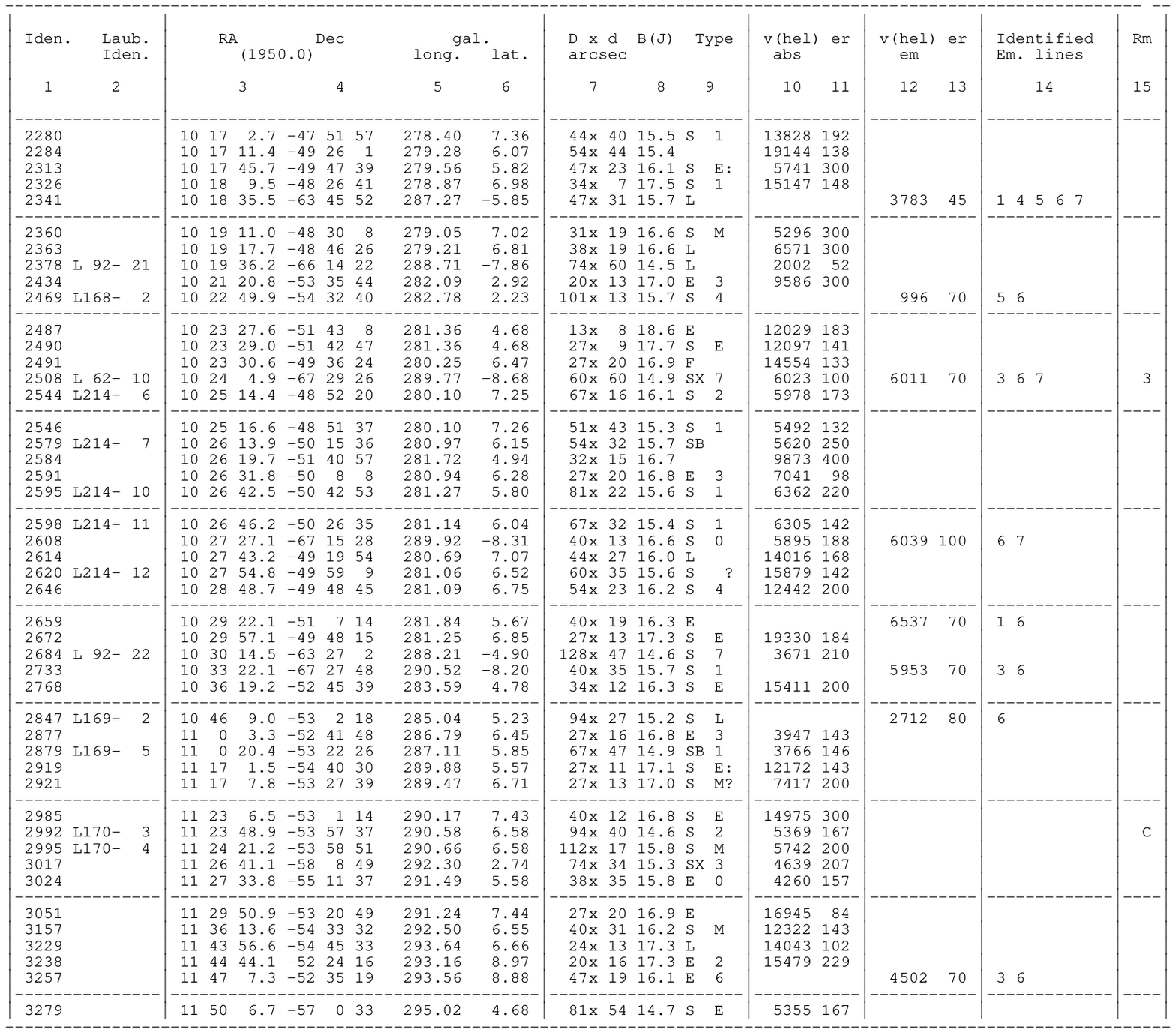}\hfil
%\vspace{15cm}
\end{table*}

The entries in Table~1 are as follows:
\begin{description}

\item [Column 1 and 2:] Identification of galaxy as given in KK94
and Lauberts Identification (Lauberts, 1982).

\item [Column 3 and 4:] Right Ascension and Declination (1950.0). The
positions were measured with  the measuring machine Optronics at the
ESO in Garching and have an accuracy of about 1 arcsec.

\item [Column 5 and 6:] Galactic longitude $\ell$
and latitude $b$.

\item [Column 7:] Large and small diameter (in arcsec). The diameters
are measured approximately to the isophote of 24.5 mag arcsec$^{-2}$ and
have a scatter of $\sigma \approx 4\arcsec$.

\item [Column 8:] Apparent magnitude B${\rm_J}$. The magnitudes are estimates
from the IIIaJ film copies of the ESO/SRC Survey based on the above given
diameters and an estimate of the average surface brightness of the galaxy.
A preliminary
analysis of this data finds a linear relation from the brightest to the
faintest galaxies (B${\rm_J} \approx 19\Mag5$) with a scatter of only
$\sigma \approx 0\Mag5$.

\item [Column 9:] Morphological type. The morphological types are coded
similarly to the precepts of the RC2 (de Vaucouleurs \etal 1976). Due to
the varying foreground extinction a homogenous and detailed type
classification could not always be accomplished and some
codes were added.
In the first column F for E/S0 was added to the normal designations
of E, L, S and I. In the fourth column the subtypes E, M and L are
introduced next to the general subtypes 0 to 9. They stand for
early spiral (S0/a-Sab), middle spiral (Sb-Sd) and late spiral or
irregular (Sdm-Im). The cruder subtypes are a direct indication of the
fewer details visible in the obscured galaxy image. The
questionmark at the end marks uncertainty of the main type, the colon
uncertainty in the subtype.

\item [Column 10 and 11:] Heliocentric velocity
and error as derived from the absorption features ($\kms$). The errors may
appear large as they are estimated external errors, and not internal errors
(see Sect. 2.1).

\item [Column 12 and 13:] Heliocentric velocity and error measured
from the emission lines (identified in column 14) when present ($\kms$).

\item [Column 14:] Identified emission lines. The codes 1-7 are explained
in the table below, where the $1^{st}$ line gives the code of the
emission lines used for the
redshift evaluation, the $2^{nd}$
line names the emission lines, and the $3^{rd}$ line lists the
corresponding wavelengths in $\AA$:\\
\begin{tabular}{ccccccc}
 1 & 2 & 3 & 4 & 5 & 6 & 7 \\
 $[$OII] & H$\gamma$ & H$\beta$ & [OIII] & [OIII] & H$\alpha$ & [NII] \\
 3727 & 4340 & 4861 & 4959 & 5007 & 6563
          & 6584
\end{tabular}

\item [Column 15:] Code for additional remarks:\\
1 -- \hspace{0.15cm} The spectrum of KK527 shows this to be a Seyfert 1
galaxy.\\
2 -- \hspace{0.15cm} The spectrum of KK1963=L92-10 was measured at different
sites and -- based on the emission lines [OII], H$\beta$,
[OIII], H$\alpha$, [NII] -- the following velocities were obtained:
$v_{hel}=134, 26, 87, 10, 35, 6, 100 \pm 58 \kms$. The velocity given in
Table~1 is the mean of all these velocities. Further details are given
below.\\
3 -- \hspace{0.15cm} Due to superposition of a star on the galaxy KK2508=L62-10
the exposures were centred slightly west of the centre of the object.\\
For some of the galaxies other redshift measurements are availabale
as well (cf. section 2.2.2):\\
A -- \hspace{0.15cm} $v_{hel}= 2916 \pm 45\kms$ by Strauss \etal 1992,
for KK1517=L126-24.\\
B -- \hspace{0.15cm} $v_{hel} = 1830\kms$ by Acker \etal 1991 for KK1853
(cf. special cases). \\
C -- \hspace{0.15cm} $v_{hel}= 5523\kms$ by Dressler 1991 for KK2992=L170-3.
\end{description}

\noindent{\bf Special cases:}
\paragraph{L166--P18} A very intriguing galaxy candidate was detected
in the galaxy search at $9^{\rm h}29^m08\fs8 -52\deg56\arcmin43\arcsec$.
It lies extremely close to the actual dust equator ($\ell=275\fdg5$,
$b=-1\fdg3$) and was one of the primary candidates for a highly
obscured nearby large galaxy. It has a very high surface brightness, is
clearly elongated ($23\arcsec$ x $15\arcsec$) with small protrusions
at the end of the major axis which turn off at opposite angles from the
main body. It looks like the most inner part of a
early type (barred?) spiral galaxy. However, the spectrum we obtained
at the SAAO clearly proved this to be a planetary nebulae -- as categorised
in Lauberts (1982) -- with its emission lines H$\beta$, [OIII],
H$\alpha$ and  [NII], and line ratios typical for PN's. The mean velocity
of this object is $v_{hel}= 7 \pm 58 \kms$.

In the course of our galaxy search, further PN-candidates have been
found. We have observed these as well and so far
discovered 2 new planetary nebulae. This is reported in
Kraan-Korteweg \etal 1994b.

\paragraph{KK1853=L92--G?02} This fairly bright elliptical galaxy at
($\ell=285\fdg9$, $b=-7\fdg7$) has been misidentified in earlier years as
a planetary nebula. In 1991, Acker \etal started a project
to get spectra of all misidentified PN. During this effort, KK1853
was confirmed to be an emission line galaxy -- as found here.

\paragraph{KK1963 = L92--G10} The morphology of this object suggests this to
be a nearby spiral galaxy. It has a patchy appearance, is quite extended
($74\arcsec$ x $74 \arcsec$) and the estimated magnitude is about
$B_J = 14\Mag3$. It lies at $\ell=286\fdg3$,
$b= -6\fdg9$ in the Milky Way, hence is seen through a non-negligible
extinction layer.  Its true dimensions will be
considerably larger and it could
well be a fairly local galaxy.
However, its identification as a galaxy is still questionnable.
In Lauberts (1982)
it was classified as an emission nebula {\em or} galaxy,
and in the Southern Galaxy Catalogue (Corwin \etal 1985) as an
irregular galaxy of type IB(s)m?.
The object has an entry in the IRAS Point Source Catalog. The fluxes
in Jy at the 4 wavebands are $f_{12}=0.42L$, $f_{25}=0.25L$, $f_{60} =
2.94$ and $f_{100} = 5.74$; L indicates that it is a lower
limit only. The colors are compatible with it being a galaxy
(\eg Meurs and Harmon, 1988, Lu \etal  1990).
But since the signature of various galactic objects are similar
to galaxies as a whole, this is not really decisive.

We did various exposures at different positions on this object and
found strong emission all over. The systemic velocity -- taken as the mean
of the emission features at the
different locations -- is
of $v = 57\kms$. The correction of this observed velocity
to the centroid of the Local Group is quite large ($\Delta = -299\kms$,
in accordance to the
precepts given in the RSA, cf. Sandage and Tammann 1981)
and the corrected recession velocity is $v_0 = -242\kms$. Based on
this velocity, the regarded object is more likely to be a galactic object
than a new member of the Local Group as its velocity is more
negative than any of the other known members of the Local Group.

\paragraph{KK2463 = L168-G02} This galaxy is the only one in the search area
with a velocity less than $1000 \kms$, namely $v=996\kms$, respectively
$v_0=689\kms$
when corrected to the centroid of the LG. It does not
lie in any known group. The nearest known group is the CenA group
at $l\approx 310\deg, b\approx 20\deg$ and $v_0 \approx 275\kms$
which definitely is too far apart on the sky as well as in velocity
space to be related.

The galaxy is an extended very LSB spiral (Sbc) at very low latitude
($\ell=282\fdg8$, $b=2\fdg2$). Based on the HI-column density (Kerr
\etal 1986) and the relation between HI and extinction as given
in Burstein and Heiles (1982) the extinction is estimated to be
at least of the order of $A_{B}\approx2\Mag5$. Considering that
the observed
properties are $101\arcsec$ x $13\arcsec$ and $B_J = 15\Mag7$, the
extinction-corrected diameters will roughly be about
a factor of 3 larger, making it an edge-on spiral of about
5 arcmin with an absorption-corrected magnitude of $B_J^0 \approx
12\Mag5$ (cf. Cameron 1990). There seem
to be a number of similar, extended LSB objects in the vicinity
of this galaxy at even lower latitudes. It is therefore not unlikely
that this galaxy is part of a nearby, previously unrecognized group
which then would not be unimportant with regard
to the dynamics in the local neighbourhood.

\subsubsection {Comparison of the SAAO-redshifts to other values}

We concentrated mainly on observing galaxies without previously
measured velocities. A comparison of our redshifts with other (published)
determinations is only possible for 3 of the galaxies in Table~1,
\ie KK1517, KK1853, KK2992. The velocity values and their sources
are given in the remarks to Table~1. A comparison between our velocities
and the measurements from the literature finds:
\begin{displaymath}
<v_{SAAO} - v_{pub}> = -106 \pm 26 \kms.
\end{displaymath}
\noindent With the exception of KK1853, the differences are well within
the errors quoted in Table~1. The offset of $-106\kms$ is quite large. It
is, however, only based on 3 independent sources, each for a different galaxy,
and therefore not really significant. This is confirmed by our
complementary observing programs: 10 of the galaxies observed at the
SAAO are in dense regions and were also observed with the multifiber
system Optopus. The offset between these observations is
\begin{displaymath}
<v_{SAAO} - v_{OPT}> = +19 \pm 75 \kms,
\end{displaymath}
\noindent hence negligible. The individual differences are within the quoted
errors. In addition, 7 of the galaxies observed here have meanwhile
been observed by us at Parkes in HI. The agreement is excellent and the
mean difference is only
\begin{displaymath}
<v_{SAAO} - v_{PKS}> = +31 \pm 20 \kms,
\end{displaymath}
\noindent with a very small dispersion.

Overall the correspondence between the radial velocities
measured at the 1.9m telescope of the SAAO and the redshifts
derived from other observing programs is very good and well within
the quoted errors.

\subsubsection{Galaxies without reliable redshifts}

For 25 of the observed galaxies no reliable redshift could be
determined. For three galaxies the bulk of the spectrum was
due to a superimposed foreground star. For three other
galaxies a tentative identifaction based solely
on the Mg absorption line at $5175\AA$ could
be made, but as no other features in
accordance with the indicated redshift could
be found, the deduced values are too uncertain.
No reliable feature could be indentified in
7 of the observed galaxies. For 12 further
galaxies the S/N was too low. Some of these galaxies
are just too faint for detection with the 1.9m telescope
even though their apparent brightness is quite bright (cf. column 8
in Table~2), but they are extended very low surface-brightness
spirals. Such galaxies are likely to be nearby spirals. We have observed
these in Parkes (Kraan-Korteweg, Henning, Schr\"oder and van Woerden,
in preparation) and the majority
of them have been detected and confirmed to
be nearby ($v< 5500 \kms$).

\begin{table*}
\caption{Galaxies observed at the SAAO without reliable redshift}
\hfil\epsfbox{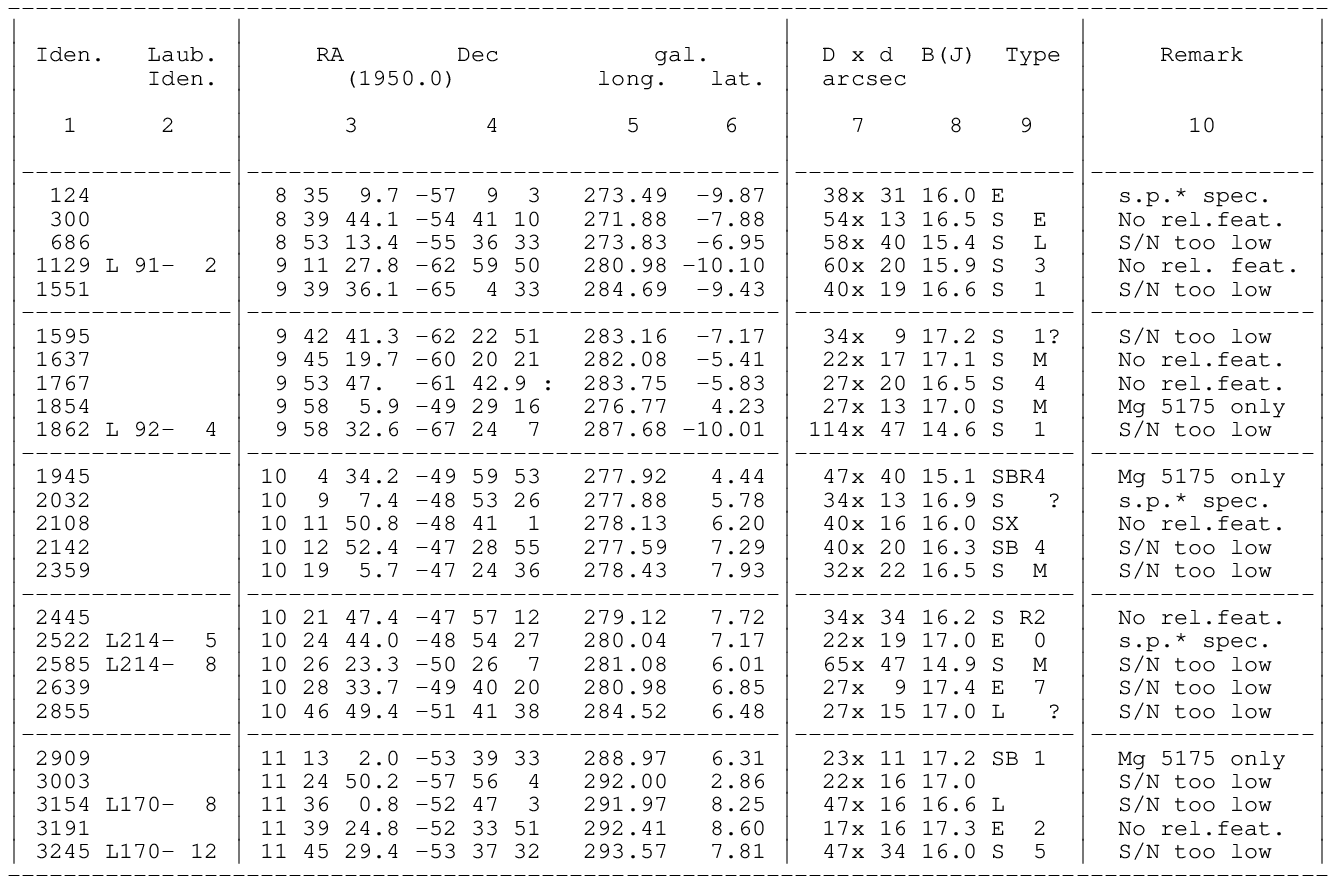}\hfil
\label{t2}
\end{table*}
The galaxies for which no reliable redshift
could be extracted are listed in Table~2. Columns
1 to 9 are the same as for Table~1 and lists names, positions and
properties of the galaxies (cf. explanations to Table~1 for further details).
The remark in column 10 describes the reason for the
non-detection.

\subsubsection{Galaxies with redshifts from the literature}
The redshifts of 31 galaxies in the ZOA in the Hydra/Antlia search area
have been determined by others. This concerns mainly bright and large
galaxies ($D>1arcmin$), plus some IRAS-galaxies within $5\deg < |b| < 10\deg$
identified by Strauss \etal 1992 for their 2Jy IRAS Redshift Survey. On
average, these are galaxies which would have entered our observing list,
if there redshift had not been measured by others. They are
relevant for the discussion of the observed sample as a whole and
these galaxies are listed in Table~3.

\begin{table*}
\caption{Galaxies with radial velocities from the literature}
\hfil\epsfbox{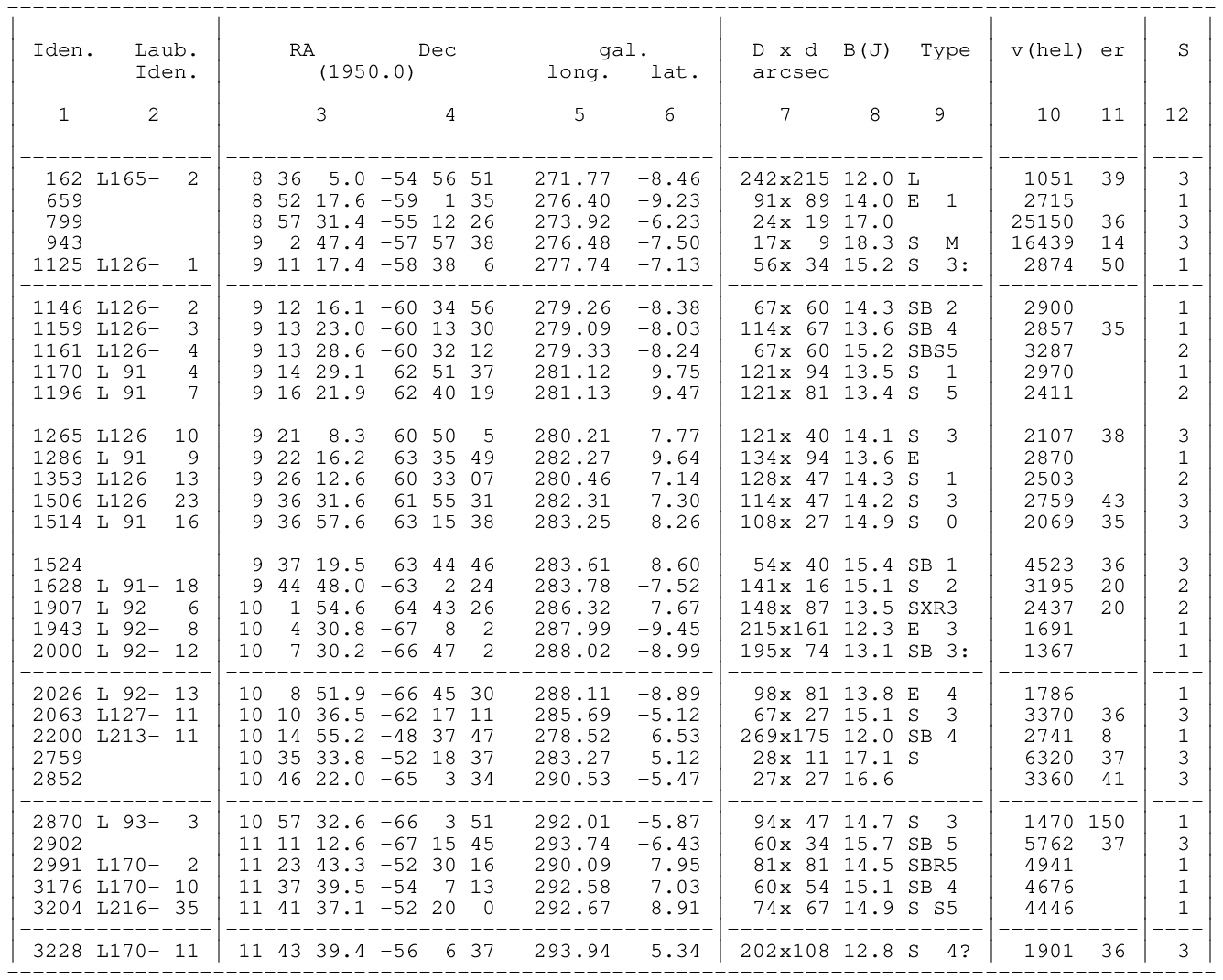}\hfil
%\vspace{10.6cm}
\label{t3}
\end{table*}
Columns 1-9 are the same as in Table~1, and details about
the entries can be found there. Column 10 and 11 list the heliocentric
velocities and errors (if given). The velocity
in column 10 has been adopted from the source identified in
column 12, where (1) stands for {\em The Southern Redshift Catalogue
and Plots}  by Fairall and Jones (1991), and (2) for
{\em The General Catalog of HI-Observations of Galaxies}
by Huchtmeier and Richter (1989). The original sources can
be looked up there. (3) are velocities measured by Strauss
\etal (1992) for the 2Jy IRAS Redshift Survey.

\subsection {Description of Observed Sample}
In Fig.~2 we have plotted in galactic coordinates
the distribution of all the galaxies in the Hydra/Antlia search region
for which a radial velocity is now available, \ie
the 115 galaxies for which we determined a redshift
at the SAAO (filled dots) as well as the 31
galaxies with redshifts from the literature (open
dots), hence 146 galaxies in total.
The 25 observed galaxies for which no
redshift could be determined are not displayed, nor the uncertain galaxy
with $v_0=-242\kms$ and the observed PN.

%\special{psfile=fig2.ps}
\begin{figure}
\hfil\epsfbox{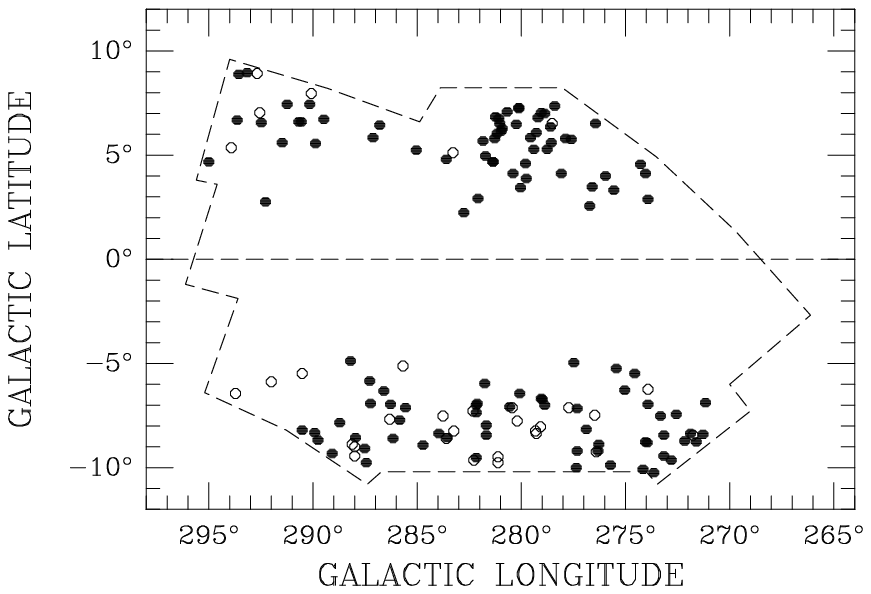}\hfil
%\vspace{6cm}
\caption{Distribution of the galaxies with radial velocities.
The Hydra/Antlia survey region is outlined. Filled dots
are objects observed by us at the SAAO; open dots are galaxies
with redshifts from other sources.}
\label{f2}
\end{figure}

As discussed in the introduction, we aimed  primarily at obtaining radial
velocities for all the brightest galaxies in the Hydra/Antlia search, or
more accurately due to the restrictions of the 1.9m telescope for all
galaxies with high central surface brightness.
At the same time, a representative coverage of the overall galaxy distribution
was attempted. Our selection criteria were softened towards the
galactic equator where a larger fraction of relatively faint and small
galaxies entered our observing program, as we tried to trace the galaxy
distribution as deep as possible into the dust equator where the
extinction is highest and the diminishing effects largest. This is partly
reflected in the distribution here.
Overall, the achieved distribution of galaxies with radial velocities
seems a fair tracer of the galaxy distribution
in the ZOA -- compared to the
distribution of all the 3279 galaxies discovered in the galaxy search
(see Fig.~1) -- with its various distinct over- and underdensities.

How representative are these galaxies
compared to the survey in general? To assess this, we
have plotted the magnitude and diameter distribution of
the galaxies with redshifts in Fig.~3. In the top panel
the number of galaxies with redshifts are
displayed, while the bottom panel shows the completeness
achieved with respect to the deep galaxy sample in the ZOA in the Hydra/Antlia
extension (KK94). The hatched areas are based on the observations at
the SAAO, the cross-hatched on others.

%\special{psfile=fig3.ps}
\begin{figure}
\hfil\epsfbox{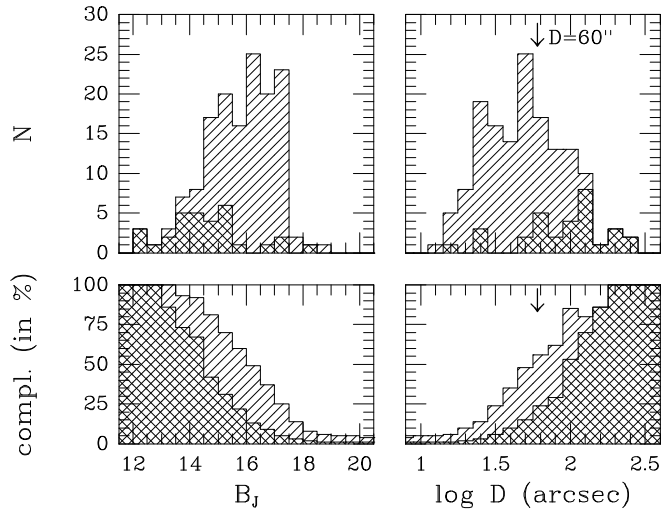}\hfil
%\vspace{6.8cm}
\caption{Magnitude and major-axis diameter distribution of galaxies with
radial velocities in the Hydra/Antlia search area. Hatched areas mark
galaxies observed at the SAAO, cross-hatched by others.
The top panel gives the number of galaxies while the bottom panel displays
the completeness in percent of the total number of galaxies in the deep
galaxy search (cf. Fig.~1).}
\label{f3}
\end{figure}

The magnitude histogram, combined with the completeness distribution,
clearly implies that we can trace the bright end of the
magnitude distribution quite well with our observations.
Although we have redshifts of 'only' 4\% of the deep Hydra/Antlia
galaxy survey, we are 92\% complete for galaxies brighter than
$B_J=14\Mag5$, a typical magnitude limit for complete
redshift surveys as, for instance, the first CFA survey (Davis \etal 1982),
and 70\% complete for $B_J=15\Mag5$, the limit for the extended CFA
survey (Huchra \etal 1994). We are even 50\% complete down to $B_J=16\Mag5$.

Although high central surface brightness, and not the apparent B$_J$-magnitude,
is the dominant condition
for observations at the SAAO, the histogram in Fig.~\ref{f3}
effectively finds a very steep magnitude cut-off at B$_J=17\Mag3$.
Above this magnitude the 1.9m telescope clearly is not sensitive enough to
obtain a reasonable spectrum from the galaxies. At the fainter
end of our magnitude distribution ($16\Mag5 \la B_J \la 17\Mag3$),
the restriction of high central surface brightness automatically
leads to the selection of very compact, face-on small galaxies,
the majority of which often are of early type and quite
distant (60\% with  $B_J>16\Mag5$ have $v>10000\kms$).

The few bright galaxies without redshifts typically are very extended
low surface brightness galaxies. This can be deduced indirectly from
the diameter distribution which is much broader,
nearly gaussian in shape: the distribution does not rise as steeply
compared to the magnitude distribution, nor does it
have a sharp cut-off for smaller galaxies. The reason is that many of the
large galaxies are spirals with an extended large (obscured) low
surface brightness disk. They are bright only (apparent magnitude)
because of their dimensions.
This is evident also from the completeness distribution: we are
'only' 62\% complete for $D>60\arcsec$ $(log D = 1.78)$,
48\% for $D>45\arcsec$ (1.65), but
still 35\% for $D>30\arcsec$ (1.48). The selection of high surface
brightness results in this relatively high fraction of small and
compact galaxies and therewith to a much
shallower decline of the detections towards smaller galaxies.

As mentioned before, the missing large low surface brightness galaxies
($D>60\arcsec$) have since been observed at Parkes
(1993 and 1994). Only few remained undetected.
It will be interesting to do follow-up observations of these few extended
low surface brightness objects to test whether these are fairly local
gas-poor dwarfs.

Altogether it can be maintained that the observed galaxies are
representative of the bright end of the luminosity distribution
of the galaxies in the Zone of Avoidance. However, it should be stressed
again
that the above parameters are uncorrected for extinction. The absorption
is of the order of $1^{mag}$ at the borders of our search area, increasing
to $5-6^{mag}$ closer to the dust equator. For higher obscuration values
the galaxies are obscured away and are not visible anymore on the IIIaJ,
nor on the infrared I sky survey plates of the ESO/SERC.

In the forthcoming
catalogue (KK94), the HI-column densities will be listed for each
galaxy. This can be used  as a rough estimate of the foreground extinction.
Together with the formalism of Cameron (1990), we can then correct
for the diminishing effects on the
diameters and magitudes of the galaxies. This correction will be quite
rough and does, of course, not account for the locally varying gas-to-dust
ratios. In this first paper on redshift measurements in the ZOA we
do not yet attempt any corrections for extinction or an assessment
of the completeness of the sample under discussion with regard
to the extinction-corrected parameters. We have currently began
a program to derive accurate extinction values
in the ZOA. A brief outline is given in Kraan-Korteweg \etal (1994a).
This will be implemented in our final analysis -- which will include
the data from all observing programs -- and discussed there in full
detail.

\section {Extragalactic structures in the ZOA and its connectivity to
structures adjacent to the Milky Way}
\subsection {The velocity distribution}

Fig.~4 shows a histogram of the redshift distribution within the
search area. The histogram contains the 115 reliable redshift values
of extragalactic objects obtained in this
observational program (hatched area) as well as the 31
galaxies in this area with velocities published in the literature
(cross-hatched), \ie it displays the velocity
distribution of the galaxies plotted in galactic coordinates in
Fig.~2. with the magnitude and diameter distribution as in Fig.~3.

%\special{psfile=fig4.ps}
\begin{figure}
\hfil\epsfxsize 8.8cm \epsfbox{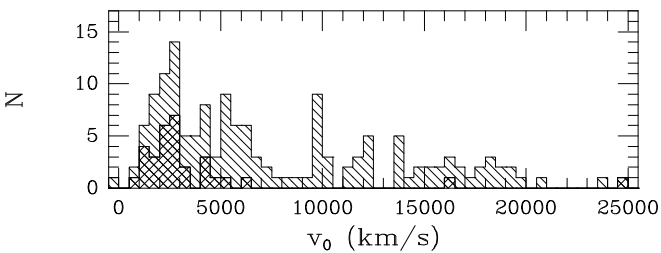}\hfil
%\vspace{3.5cm}
\caption{Velocity histogram of the galaxies in the search area
in the Hydra/Antlia extension in the ZOA. Hatched areas are velocities
measured by us; cross-hatched by others.}
\label{f4}
\end{figure}

Of the 146 galaxies in total, with redshift measurements
there are only three with $v_0 > 20000\kms$, while over two thirds (N=102)
have $v_0 < 10000\kms$.
Thus we can expect this survey to be useful for tracing structures closer
than $10000 \kms$, with some indication of structures out to $20000 \kms$.

In the nearby velocity range ($v_0 < 10000\kms$), the peaks at 2500, 5000
and $9500\kms$ stand out quite prominently. Are they significant? Are these
the first indications of extragalactic structures unveiled in the
Southern Milky Way plane?
Due to the extreme and locally varying extinction effects in our
galaxy sample it is difficult to quantify the significance of these
distinct peaks, \ie to compare the velocity distribution obtained here
with the {\em expected} velocity distribution
for a homogenously distributed galaxy sample with well defined
selection criteria, such as a diameter, magnitude or velocity
completeness limit. Our velocity data
is -- as discussed in the previous section -- almost complete
for $B_J \le 14\Mag5$ and decreases gradually in
completeness to $\approx$ 25\% at $B_J \approx 17\Mag5$.
It is quite clear, however,
that this distribution is not the result of a homeogenous distribution of
galaxies. The peaks in the velocity histogram at 2500, 5000 and 9500$\kms$
are too striking and surely originate from individual extragalactic
structures such as clusters, filaments and possibly voids. We will
therefore now try to localise the features
suggested in the histogram here in 3-dimensional space. This can be
visualized best by studying the distribution of the observed galaxies
in various projections.

\subsection{3-D distribution of the galaxies in the search area
and its surroundings}

Our sample volume can be projected or "sliced" in various ways,
to examine
the distribution of the data. Experience has shown that particular
structures often show up best in certain favoured directions and not in others.
Aside from the data of the present survey, we will also extend the plots
to include surrounding regions (similarly as in Fig.~1). The data from
the areas adjacent to our survey are taken again from the compilation
of the Southern Redshift Catalogue (SRC) by Fairall and
Jones (1991) and the 2Jy IRAS Redshift Survey (Strauss \etal 1992).
They are combined with the velocities obtained here
to reveal the connectivity with known structures and/or disclose new ones.

In studying the 3-D distributions in and around our survey
region, the reader should be aware of an unavoidable
selection effect: the combined data is uncontrolled in
magnitude. The true apparent magnitudes of our surveyed
galaxies are, as discussed above, very uncertain, due to the
irregularities of the foreground extinction. The SRC data is
a compilation of published redshifts from all available sources,
and does not provide a statistically-controlled
sample. Experience has shown though that it contains most
of the galaxies with B$_J < 15\Mag0$ and $D > 1\arcmin$ north of our
survey region. South of our sample area, the coverage is
less complete, since the sky towards the South Celestial
Pole has received less attention. Due to the obscuration of
the ZOA, and therefore the fact that most extragalactic studies
concentrate on high-latitude region, the strip
$10\deg < |b| < 20\deg$ is strongly undersampled in the SRC,
while very few SRC redshifts are,
of course, found with $|b| < 10\deg$. This implies that there
is as yet hardly any data for the ZOA {\em on either side of our survey
area}.

The absence of magnitude control implies that the density of
galaxies projected onto the sky generally does not give a reflection
of the true density (for the survey region itself, a detailed
discussion of the achieved coverage compared to the deep galaxy
search has been presented above).
Distinct redshift peaks, however, are completely independent
of these selection effects. Extinction effects and
variations in limiting magnitude only affect the general
decline in galaxy numbers with increasing redshift. They can
never preselect towards a particular redshift
peak.

The accompanying plots should therfore be interpreted as follows:
an elongated feature (a concentration of galaxies in
redshift space) that runs radially with the line of sight is
suspect. It could be real (a cluster or a filament running outwards),
but it could also have been
created by a hole in the extinction (in the ZOA), or --
in the case of the SRC -- a favoured small survey region. An
elongated feature that runs across the radial direction has,
to be real, however, since it is created by a sequence of
distinct redshift peaks over different directions in the
sky. A similar argument can be applied to voids (the
opposite of a redshift peak): voids have to be real if they
are enclosed, on their far side, by a significant number of
galaxies. It is possible to establish a statistical measure
by means of a technique given in Kauffmann and Fairall
(1991), but this is a major undertaking and outside the scope
of the present paper.

\subsubsection {Sky projections}
We begin with sky projections in galactic coordinates. The
distributions are {\em sliced} in redshift intervals of
$\Delta v = 2000 \kms$ at progressively larger distances,
starting at $v = 1000\kms$ in the top left panel, moving downwards
and ending at $v = 17000\kms$ at the bottom right panel. Within
the individual panels the different
symbols give further distance information: the filled dots
outline the galaxies within the lower $\Delta=1000\kms$ velocity interval
and the crosses the higher $\Delta v$-range.
The first slice ($v < 1000\kms$) is not illustrated because it contains
only one galaxy. This galaxy (at the surprisingly
low latitude of $b=2.23\deg$) is discussed in detail under special
cases (KK2463).

%\special{psfile=fig5.ps}
\begin{figure*}
\hfil\epsfysize 21cm \epsfbox{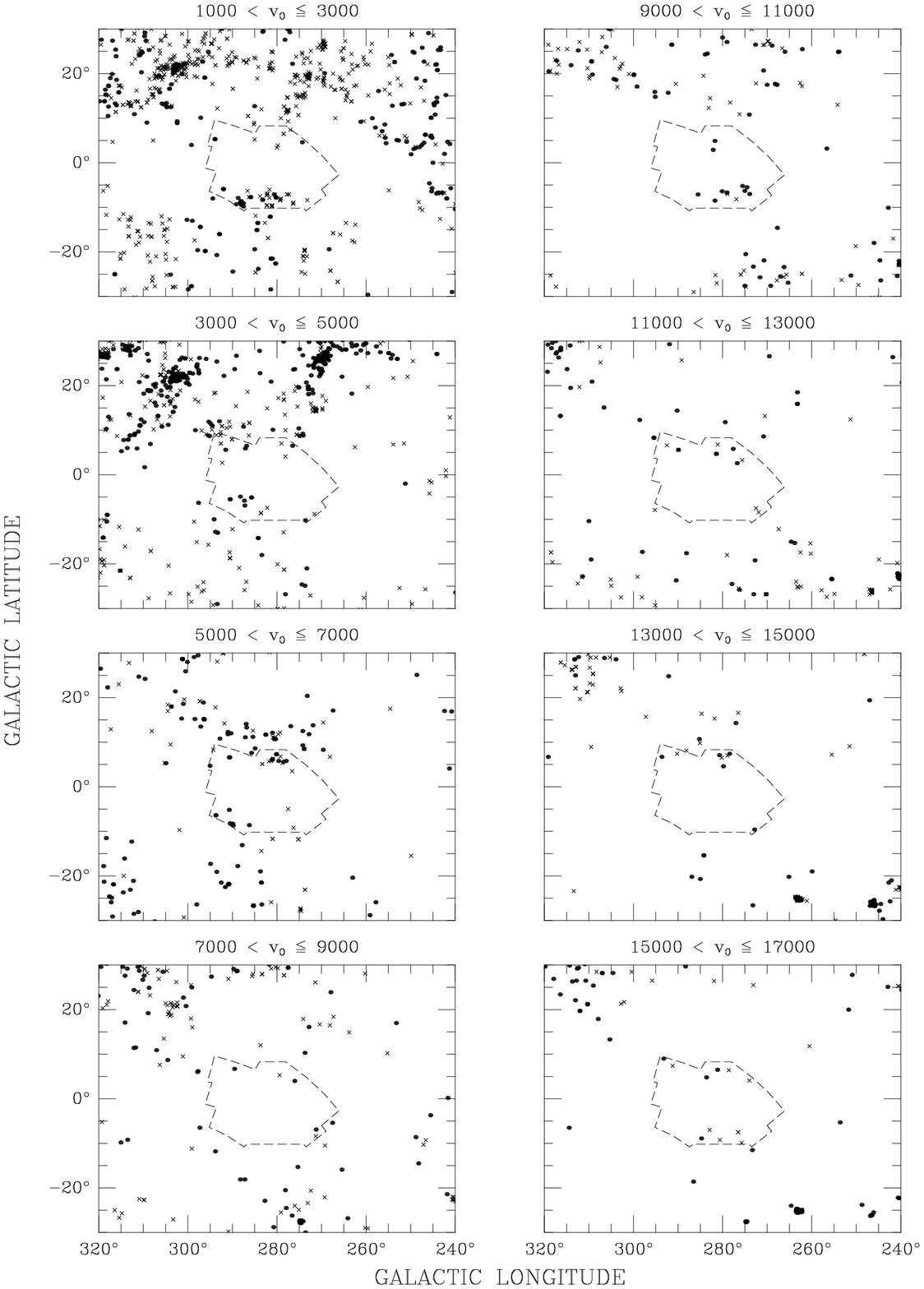}\hfil
%\vspace{22cm}
\caption{Sky projections in galactic coordinates for redshift intervals
of $\Delta v=2000\kms$. Within the panels the redshifts are subdivided
into intervals of $\Delta v=1000\kms$: filled dots mark the nearer
redshift interval (\eg $1000<v_0\le2000\kms$ in the top-left panel),
crosses the more distant interval ($2000<v_0\le 3000\kms$ in same panel).
The skyplots increase in velocity-distance from the top-left panel to
the bottom-right panel as marked above each panel.
The area of our investigation is outlined.}
\label{f5}
\end{figure*}

The first plot (top left) in Fig.~5 clearly shows the Antlia
extension as an obvious linear structure descending from top
centre of the diagram towards our survey area, so to
intersect the Galactic Plane at an angle of about 70$\degr$.
This panel shows quite clearly that the Hydra and Antlia clusters are
not isolated clusters in space but form part of a larger
structure -- as suspected -- that crosses the plane and
continues with a concentration of galaxies within our survey area south
of the plane. The linear filament is less pronounced
there; it seems to break up into two filaments.

In the second plot ($3000 < v_0 < 5000\kms$) the filamentary
continuation is hardly visible anymore within our search area.
However, due to the velocity dispersion in the cluster itself,
Antlia is still distinct, while the Hydra I
cluster, and also part of the Centaurus cluster to the left of it,
clearly are the dominant features in this panel. A more detailed
investigation of the velocity distribution manifests that
the Hydra filament starts at
a velocity of about $2500\kms$ in the galaxy concentration
below the galactic plane, flowing outwards in redshift space
to Antlia at about 2800$\kms$ and then to Hydra at $\approx 3300\kms$.
It is similar to the Puppis filament visible in the top panel,
which starts in the GP ($\ell\approx  245\deg$) at 1500$\kms$ and
leads to the Antlia cluster as well, from where it possibly continues
to the Centaurus cluster.

As discussed above, the data below the
southern Milky Way is sparse, but it is not inconceivable that
this filamentary Hydra/Antlia extension does continue even further and
connects with the Fornax cluster. This link is quite pronounced
in the distribution of the extragalactic light (cf. Fig.~2 in
Lynden-Bell \& Lahav 1988) and has been recognised independently
by Mitra (1989). He called it the Southern Supercluster and
determined its mean velocity at 1800$\kms$.

A second
wall, perpendicular to the first (\ie at about $30\degr$ to the
plane and mainly north of it), appears in the second plot
and possibly persists through to the 5th or 6th plots; it is
prominent in the third plot ($5000 < v_0 < 7000\kms$) and its
intersection with the first wall there coincides with the
Vela overdensity ($\ell\approx280\deg, b\approx+6\deg$).

A surprising result of our spectroscopic data is that the Vela
overdensity,
which lies exactly in the Antlia extension, is not -- as expected --
part of this general structure. It is prominent in the third plot,
and accounts for the redshift peak at 5000-6000$\kms$ in the velocity
histogram. The distribution of the other galaxies in this panel is
quite surprising as well: although the data coverage is minimal
adjacent to our survey region, the majority of the galaxies are found
next to this boundary while the remaining region is
nearly devoid of galaxies. In the analysis of the redshift slices
(see the middle right-hand panel of Fig.~6 and further discussions
there) this structure stands out
even more conspicuously. It seems a shallow but large-scale
overdensity.

Beyond 9000$\kms$, a concentration of galaxies south of the
plane accounts for the histogram peak at 9500$\kms$. The data
beyond is sparse, but all four right-hand plots in Fig. 5
show a top-left to bottom-right tendency, with support from
the new data in our survey area. The concentration that
occurs in the top-left corner is the Shapley region or
'Alpha' region, thought by some (\eg Vettolani \etal 1990) to
be the greatest overdensity within 20000$\kms$. The
concentration in the bottom-right corner (in
the 5th to 7th plots) are the Horologium clusters. The Horologium
and the Shapley regions dominate the
$10000-20000\kms$ velocity range.

Note that the data in our sample volume shows a substantial
number of galaxies in this redshift range -- even more so than
outside the Shapley and Horologium regions in the other slices,
and despite the fact that our sample volume is lighter sampled
than the adjacent regions.
Are the Shapley and Horologium regions {\em one major structure
bisected by the Milky Way}? Considering the claims of
the significance of the Shapley region, the revelation that it is but
a part of an even larger structure is obviously extremely important.

This feature may represent a massive overdensity -- its coincidence
with the microwave dipole a major gravitational influence
on local structures. However, this will only be the
case if this structure is singular in its existence. Until we
know the distribution of galaxies over the whole sky out to distances
beyond this large wall (or sheet), we cannot assess whether
this structure is exceptional and whether the gravitational perturbation
of its mass is not counterbalanced by similiar structures on opposite
parts of the sky. Whatever the case, the size of this structure --
if confirmed to be one coherent wall or sheet -- will have
important cosmological implications: covering about $100\degr$
on the sky at a velocity of about $\sim15000\kms$ it extends over
$\sim 250$ h$^{-1}$ Mpc. This would be the largest structure
recognised in the Universe to date, nearly twice the size of the
Great Wall.

Obviously, the scarcity of the data at this redshift
makes this finding somewhat tentative, but with the addition of our
fiber-optic data (in preparation) which will concentrate on the
densest areas in the galaxy distribution and probe much deeper
($B_J \approx 19.0 - 19\Mag5$) this important
structure will be traced in detail. So far, preliminary results from
the multifiber spectroscopy indeed support the existence
of this large wall (cf. Kraan-Korteweg \etal 1993, 1994a).

\subsubsection {Redshift slices}
We now examine the distribution in conventional 'pie-shaped' redshift
slices (wedge diagrams). Fig. 6 shows sequential slices in latitude,
\ie above the ZOA the ($+15\deg < b \le +45\deg$), in the ZOA
($-15\deg \le b \le +15\deg$) and below
the ZOA ($-45\deg < b \le -15\deg$). The panels on the left go out to
$v_0 < 5000\kms$ -- to trace the Hydat/Antlia extension, the ones on
the righthand side out to $v_0 < 10000\kms$ -- to outline the Vela
overdensity. A slightly larger area of sky is displayed compared
to Fig.~5 (the longitude range is $230\deg < \ell <
330\deg$). The new data occurs in the central slices in which the
broken lines delineates the area investigated here. Filled dots are
measurements from the SAAO, crosses from the literature.
The slice represent the same features as identified
in Fig. 5, but they are now seen
at right angles to the previous view.

%\special{psfile=fig6.ps}
\begin{figure*}
\hfil\epsfxsize 16cm \epsfbox{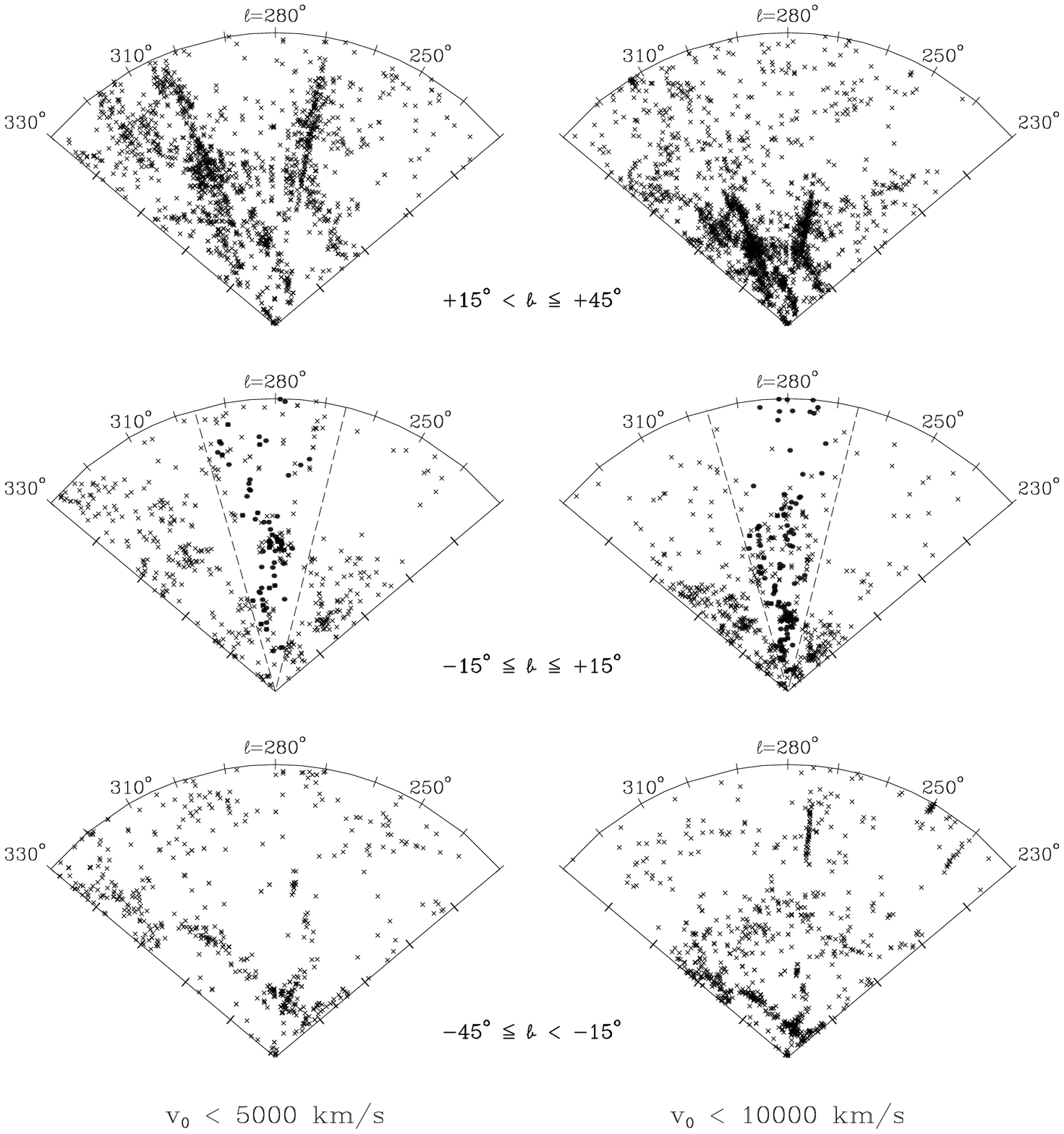}\hfil
%\vspace{19cm}
\caption{Redshift slices out to $v_0 < 5000\kms$ (left panel) and
$v_0 < 10000\kms$ (right panel) for the longitude range $230\deg < \ell <
330\deg$. The top panels display the structures above
the GP ($+15\deg < b \le +45\deg$) the middle panel in the GP
($-15\deg \le b \le +15\deg$) and the bottom panel the structures below
the GP ($-45\deg < b \le -15\deg$). The dashed lines in the middle panel
delimits the survey area. Filled dots are measurements from
the SAAO, crosses from the literature.}
\label{f6}
\end{figure*}

The uppermost
slices (Northern Galactic Hemisphere) show the dominant
Centaurus (left) and Antlia plus Hydra I (right) clusters, together with
the general concentration of the Centaurus supercluster. The
data in the lower slices is sparser, since sampling south of
the Galactic Plane and close to the South Celestial Pole is
much thinner.

At $v_0 < 3000\kms$, the new data (filled dots in the middle panel) show
a sort of filamentary
structure, previously referred to as the 'Hydra-Antlia'
wall, but which now shows a concentration at $v_0 = 2200\kms$
(the earlier sky plots show this to lie south of the Galactic Plane).
This structure represents a continuation of the Hydra-Antlia
supercluster downwards through the Galactic plane, and also
a connection to the Fornax Wall, in the lowermost plots.

Most of the other new data fill in the region $3000 < v_0
< 6000 \kms$, but the points are spread more uniformly (between the
broken lines). This is the second wall, described in
relation to Fig. 5, but now seen almost flat on - hence the
lack of discernable structure.

The galaxy distribution falls off abruptly beyond
$v_0=7000\kms$ (\cf middle right plot), creating an
underdense region up to the perimeter of the
diagram (i.e. towards $10000\kms$). Very likely
this is caused by the presence of one or more voids.
The fact that we find quite a number of galaxies
in our survey beyond $10000\kms$ (\cf also Kraan-Korteweg \etal
1993, 1994a) emphasizes that the boundary
just beyond $7000\kms$ is real and not an artifact.

In this projection the shallow galaxy overdensity around
($\ell,b,V$)$\approx$($280\deg,+6\deg,6000\kms$)
-- which was first recognised by Kraan-Korteweg and Woudt (1994a) --
is even more pronounced. In addition to the points displayed in
the middle right plot, it hosts the cluster S0639 (not shown here)
at ($\ell,b,V$)=($280\deg,+11\deg,\sim6000\kms$) as well. This
dense cluster is presently being analysed by Stein (1994).
Independent support for the existence of the Vela overdensity, as well
as its importance, is given by Hoffman (1994) who applied his density
reconstruction algorithm based upon constrained realizations of the
density field using Wiener filtering to IRAS-galaxies from the 2Jy Redshift
Survey. He predicts a massive overdensity very close to this location
and velocity ($285\deg,+5\deg,\sim6000\kms$).
It is embedded in a density contour,
which contains the Hydra/Antlia/Centaurus overdensity at about
3000$\kms$, the Great Attractor density peak at about 4500$\kms$
{\em and} the Vela overdensity at 6000$\kms$. The latter is comparable in
size and strength to the Hydra/Antlia/Centaurus peak.
Earlier reference to the existence of this structure has been given
by Saunders \etal 1991, who predicted a supercluster (S4) at this location
and distance ($281\deg,+0\deg,6120\kms$) -- although it was only a
$3\sigma$ density peak with the data available at that time. The close
correspondence of these density peaks indicate that they are consistent
with one another.

A conspicuous wall runs diagonally
downward in the last two slices. This feature was first revealed
in wedge diagrams Kraan-Korteweg (1992b)
and was previously unrecognised in plots of equatorial
coordinates. Projection in rectangular coordinates (Fairall
\etal 1994) shows it to be an extension of the 'Fornax
Wall'.

\section {Conclusions}
This paper is the first of a series with the aim
of tracing the 3-dimensional galaxy distribution in the Southern Milky Way
as close as possible to the dust equator. Our basis for these investigations
are deep optical galaxy searches done by one of us on sky survey plates.
We have presented here 115 new redshifts from
optical spectroscopy obtained with the 1.9m telescope of the SAAO.

The SAAO-observations are well suited
for observations of the bright end of the luminosity function
($B_J <17\Mag3$) of the deep galaxy search. They are
furthermore selected
to give homogenous coverage over the whole search area. The results
demonstrate that we can trace structures out to 10000 $\kms$
with indications of structures out to 25000 $\kms$.
The 115 redshifts have been combined with other published velocity data
in our search area. The most interesting features uncovered are:
\begin{itemize}
\item The earlier suspected Hydra/Antlia extension across the Milky
Way could be confirmed. It can be traced from the Hydra I cluster at
$\ell=270\deg,b=28\deg, v\sim3300\kms$ towards the here unveiled
group at $280\deg,-7\deg,\sim2500\kms$.
It seems more a filamentary structure, consisting of spiral-rich
groups and clusters, rather than a supercluster.
The Hydra/Antlia filament will be bring the dipole direction as
derived from accumulation of the gravitational forces of galaxies
in better agreement to the CMB measurement.
It is possible that this filament continues and extends to the Fornax
cluster ($240\deg,-57\deg,\sim1450\kms$).
\item The prominent galaxy overdensity in Vela is part of previously
unrecognised shallow, large-scale, overdensity centered on $\sim
6000\kms$. The independent predictions of a supercluster in the ZOA
at this positon
and distance by Hoffman (1994) and Saunders \etal (1991) indicate
that it could be quite a massive.
\item The distribution of the distant galaxies seem to bridge
the Shapley concentration and the Horologium clusters. If substantiated
with our forthcoming data, this wall/sheetlike feature of
about 250 h$^{-1}$Mpc will be the largest coherent structure
detected in the Universe, its large size difficult to reconcile with
current cosmological models.
\end{itemize}

As this is the first paper on our redshift observations, discussions
on the achieved completeness of the observed sample - which is not
straightforward to assess through the murk of the Milky Way plane --
has not been attempted.

\acknowledgements
{The authors would like to thank the night assistants Francois van Wyk
and Frans Marang as well as the staff at the SAAO for their hospitality.
The research by RCKK has been made possible by a fellowship of the Royal
Netherlands Academy of Arts and Sciences. Support was also provided by
CNRS through the Cosmology GDR program.}

\end{document}